\documentclass[doublespacing]{elsart}

\usepackage{graphicx}
\usepackage{amssymb}

\begin{document}
\begin{frontmatter}


\journal{SCES'2001: Version 1}


\title{Fulde-Ferrell-Larkin-Ovchinnikov state in $d$-wave
superconductors }

%
%
%
%
%
%

\author[us]{K. Maki}
\author[ko]{H. Won\corauthref{hwadd}\thanksref{hw}
}

%

\address[us]{Department of Physics and Astronomy,
 University of Southern California, Los Angeles, CA 90089-0484,
USA}
\address[ko]{Department of Physics, Hallym University
Chunchon 200-702, South Korea}

%
%
%
%

\thanks[hw]{HW acknowledges the support from the
Korean Science and Engineering Foundation(KOSEF) through the Grant
No. 1999-2-114-005-5.}
%
%
%
%

\corauth[hwadd]{Corresponding Author: Dept. of Physics, Hallym
University, Chunchon 200-702, South Korea.
Phone:(+82)-(33)240-1417, Fax:(+82)-(33) 254-2873,
Email:hkwon@sun.hallym.ac.kr}


\begin{abstract}

The Fulde-Ferrell-Larkin-Ovchinnikov (FFLO) state can be
found in layered $d$-wave
superconductors such as high $T_c$ cuprates and $\kappa$-(ET)$_2$
salts in a planar magnetic field. We show the superconducting
order parameter forms two dimensional crystalline lattice in the
FFLO state for $d$-wave superconductors.
Also the quasiparticle  density of states and the
thermodynamics of FFLO state are constructed. Therefore STM or NMR
will provide a definitive test for the existence of the FFLO state.

\end{abstract}

%
%



\end{frontmatter}

%
%
%
%
%


When the Pauli paramagnetism or the Zeeman term dominates the
orbital effect, the superconducting state in a magnetic field
enters in a new state where the superconducting order
parameter varies periodically in space\cite{1,2}.
However the realization of this condition is extremely difficult
in the classical $s$-wave superconductors.
First of all the system has to be in the superclean limit
where the quasiparticle mean free path $l$ is much longer than
the coherence length $\xi$.
Second the system has to have the Ginzberg Landau parameter $\kappa$
much larger than the unity.
Indeed these two conditions appear to be met readily in
$d$-wave superconductors like  high $T_c$ cuparate superconductors
and organic superconductors  like $\kappa$-(ET)$_2$ salts and
$\lambda$-(ET)$_2$ salts \cite{3}.
\par
Earlier we have shown that the region of the
Fulde-Ferrell-Larkin-Ovchinnikov (FFLO) state in $d$-wave superconductors
is much
more extended than the corresponding one in $s$-wave
superconductors \cite{4}.
In  the mean time  a possible observation of the FFLO state in
$\kappa$-(BEDT-TTF)$_2$Cu(NCS)$_2$
in a planar magnetic field has been reported \cite{3}.
Indeed in the pure Pauli limiting case we expect the upper
critical field $H_{c2}(t,\theta)$ is independent of the
direction of the magnetic field as long as $\bf{H}$
lies in the conducting plane \cite{4}.
\par
The object of this paper is to extend the earlier analysis to
construct the free energy associated with FFLO state.
Indeed  in the limit of $T\rightarrow 0$,
following Larkin and Ovchinnikov we can show that the two
dimensional periodic solution
$\Delta({\bf r}) \sim \cos(qx) +\cos(qy)$ is the most stable.
Also we obtain simple expression of the quasiparticle density
of states, the specific heat, the magnetic susceptibility and
$T_1^{-1}$ in NMR which should be readily accessible
experimentally.

In Fig.1 we show the magnetic phase diagram of $d$-wave superconductors.
 As is
readily seen, the direction of $\bf{q}$ in the most stable FFLO state
switches from $ ${\bf q}$ \parallel (1,0,0)$,  $(0,1,0)$, etc. (the direction
of 4  maximum gaps) in the
limit of $T=0$ to $ ${\bf q}$ \parallel (1,1,0) $,
$(1,-1,0)$, etc.( the direction of 4 nodes) for $T > 0.05$ $T_c$.
In the  first region(i.e.
 ${\bf q}$ is the direction of one of 4 maximum gaps)
$q=2\frac{h}{v_F}$ at $T=0$,
where $h=\mu_B H$,
while in the second region
(i.e. $ ${\bf q}$ \parallel (1,1,0)$, etc.)
$q \simeq 2.42 \frac{h}{v_F} $.
At $T= 0$ K the upper critical field is given by
$h_0=\frac{\pi}{\gamma}
T_c \exp(\frac14\cos(4\theta))$
with $q=2\frac{h}{v_F}$ and $\theta=0$,
where $\theta$ is the angle between
${\bf q}$ and the $a$-axis.
Here $\gamma$ is the Euler constant.
\par
Then following\cite{2}  we construct the equation
for $\Delta({\bf r})$, assuming that
$ \Delta({\bf r}) = \sum_m \Delta_m e^{\imath {\bf q}_m \cdot {\bf r}}$,
as
\begin{eqnarray}\label{1}
\Delta^\ast_n
& =& \sum_m \{ (2 -\delta_{mn} |) \Delta_m|^2
\Delta_n \hat{J}(q_n,q_m)
+ \nonumber\\
& & (1 - \delta_{mn} -\delta_{-m n}) \Delta^\ast_m
\Delta^\ast_{-m} \Delta_n \tilde{J(q_n,q_m)} \}
\end{eqnarray}
where
$J(q_n,q_m)$ and $\tilde{J}(q_n,q_m)$ are defined in the same way as in \cite{2}.
But here we generalized their treatment to  the $d$-wave superconductivity.
Both $J(q_n,q_m)$ and $\tilde{J}(q_n,q_m)$ depend on only angle between ${\bf q}_m$ and ${\bf q}_n$. We have   $h^2 J(0) =-15.5$,
$h^2 J(\pi)=3/2$, $h^2 J(\pi/2)=3.0$ and
$ h^2\tilde{J}(\pi/2)=5/6$

Then we can solve for different forms for the lattice. Here are; \\
(a) $\Delta({\bf r}) \sim e^{\imath {\bf q} \cdot {\bf r}} $, we have
$|\Delta|^2 = -0.065 h_0(h_0 -h)$ \\
(b) $\Delta({\bf r}) \sim \cos(qx)$, we have
 $|\Delta|^2 = -0.08 h_0(h_0 -h)$ \\
(c)$\Delta({\bf r}) \sim \cos(qx) + \cos(qy)$, we have
$|\Delta|^2 = \frac67 h_0(h_0 -h)$ \\
Therefore the solution (c) is the only one which is stable in the layered $d$-wave superconductors.
Limiting ourselves to $d$-wave superconductors
in a planar magnetic filed, we conclude
the two dimensional periodic solution, i.e. 2 dimensional square lattice,
is the
most stable one in the limit
$T \rightarrow 0$ K.
This is in a sharp contrast to the case of 3D $s$-wave
superconductors where a stripe-like state is favored \cite{2}.

The  quasiparticle density of states in the vicinity
of $h=h_0$ is given by

\begin{equation}
  N(E)/N_0 \simeq 1+ \frac{\Delta^2}{4} \sum_{\pm}
  \big < \frac{\cos^2(2\phi)}{[E \pm h + \frac{vq}{2} \cos\phi]^2} \big>  = 1 +
  \frac{\Delta^2}{2 h^2} J(\frac{E}{h},\frac{vq}{2h})
  \end{equation}
and
\begin{eqnarray}
J(\varepsilon,p) &=&\frac12 \sum_{\pm}\mathrm{Re}\big\{
\frac{|\varepsilon\pm 1|}{[(\varepsilon\pm 1)^2-p^2]^{3/2}}
+ \frac{4}{p^4}\big[(3(\varepsilon\pm 1)^2- \frac{p^2}{2})
\nonumber\\
&  & -(3(\varepsilon\pm 1)^2- 2p^2)
\frac{|\varepsilon\pm 1|}{\sqrt{(\varepsilon\pm 1)^2 -p^2}} \big] \big\}
\end{eqnarray}
and
\begin{equation}
J(0,p) = \mathrm{Re}\frac{1}{(1-p^2)^{3/2}}  +
\frac{4}{p^4} (3 -\frac{p^2}{2} -(3-2p^2) \mathrm{Re}
\frac{1}{\sqrt{1-p^2}} )
\end{equation}

In Fig.2 we show the density of state for $p(=vq/2h)=1$, 1.1 and 1.21.
As readily seen from Eq.(2) N(E) diverges for $E=(p-1)h$ and
$E=(p+1)h$ where $p=vq/2h$ for $p \ge 1$.
When $p=1$, N(E) diverges at $E =0$ and $E= \pm 2h$.
On the other hand when $p > 1$,
the  density of states at Fermi energy  is given by
\begin{equation}
 N(0)/N_0 =1 + 2(\frac{\Delta }{h})^2p^{-4}(3-\frac{p^2}{2})
\end{equation}
$N(0)$  can be accessible   through the spin susceptibility at $T=0$ K,
 $\chi/\chi_n = N(0)/N_0$, or the nuclear spin lattice relaxation rate
 at $T=0$ K, $T^{-1}_1/T^{-1}_{1n} =(N(0)/N_0)^2$ etc.
 It is shown in the inset of Fig.2.
 Also for $p>1$, $N(E)$ has two peak at
 $E=(p-1)h$ and $E=(p+1)h$.
 These peaks should be detectable  by STM for example.
 Therefore the presence of 2 peaks provides  a clear sign of FFLO
 in $d$-wave superconductors.

%
%
%
%

 \begin{figure}
     \centering
   \includegraphics{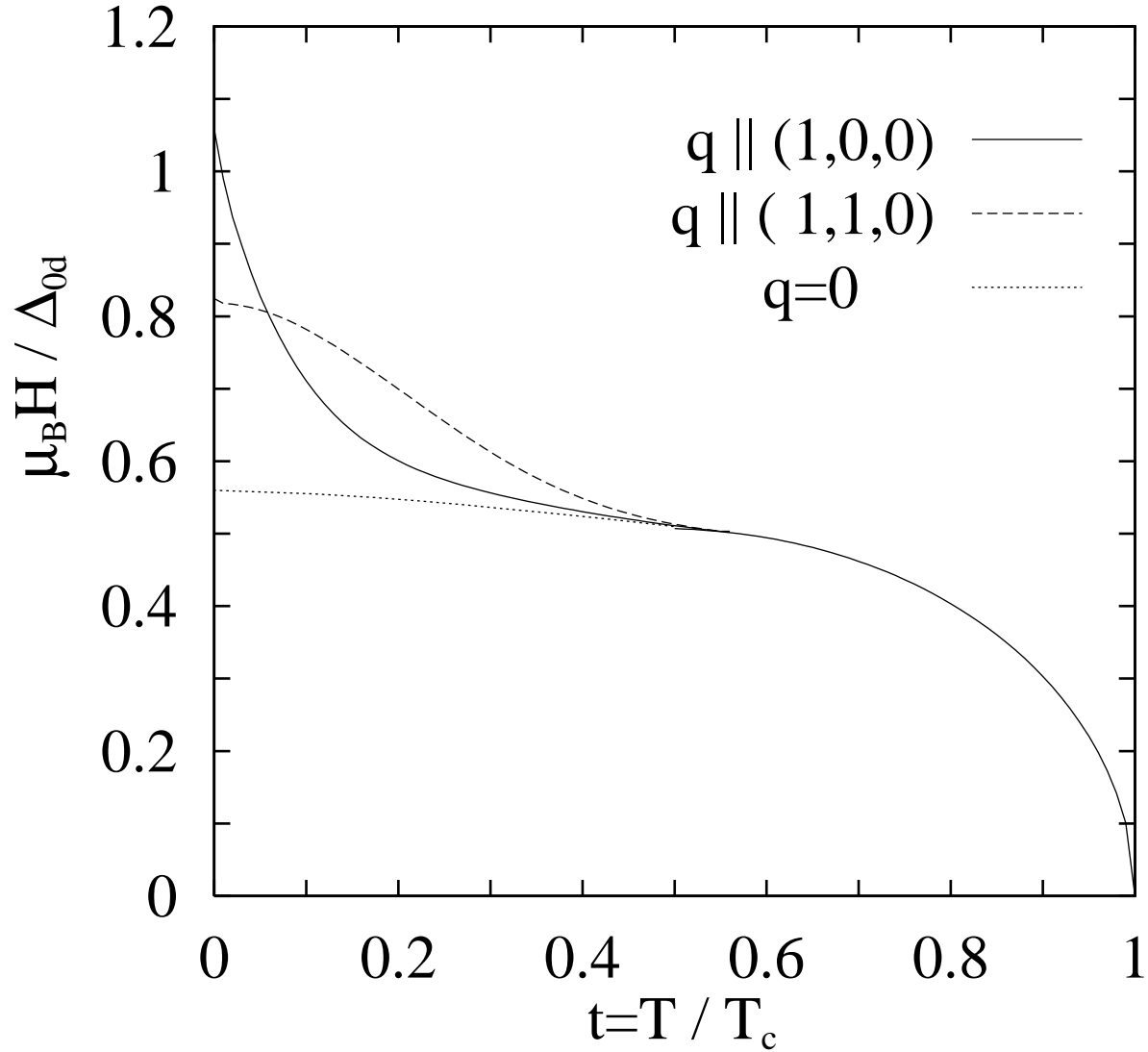}
    \caption{The phase diagram of FFLO state in $d$-wave
    superconductor.
    The solid line describes the critical  magnetic field for
    ${\bf q} \parallel $ (1,0,0), while the broken line for
    ${\bf q} \parallel $ (1,1,0).
    The dotted line is the 1st order transition line for the uniform state.
     $\Delta_{0d}$ is the $d$-wave superconducting order parameter at $T=0$. }
 \end{figure}

 \begin{figure}
     \centering
   \includegraphics{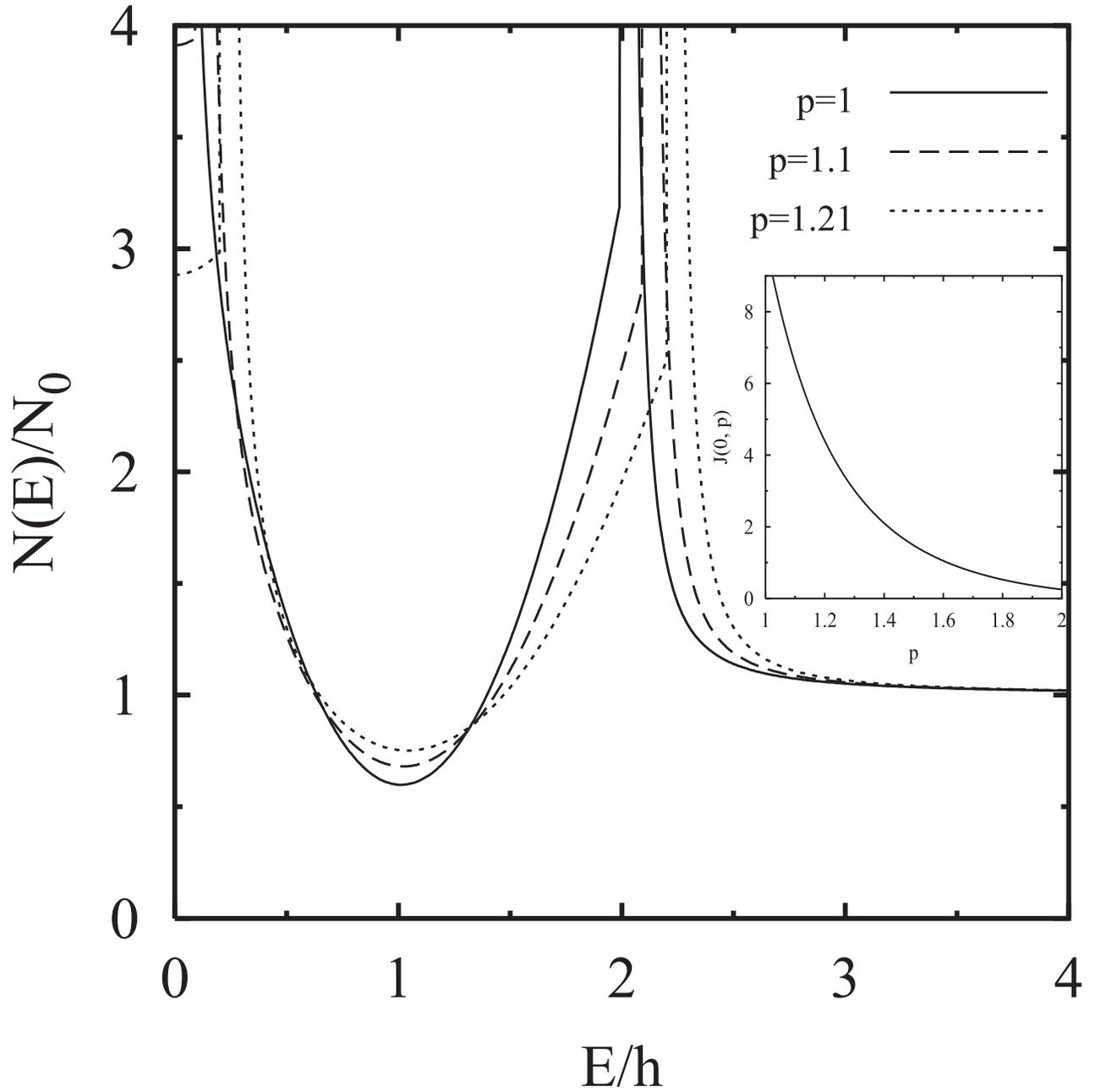}
    \caption{The density of state in Eq.(2) for $\frac{vq}{2h}
    (\equiv p) =1$, 1.1 and 1.21 is shown.
    $J(0,p)=2h^2/\Delta^2(N(0)/N_0 -1)$ is shown in the inset.
        }
 \end{figure}
%
%
%
%
%


\end{document}